\documentclass[12pt]{article}
\usepackage{amssymb}
\usepackage{amsmath}

\usepackage{amstext}
\usepackage{graphicx,epsfig}
\usepackage{epsfig}
\usepackage{verbatim} 
\usepackage{fancyhdr}
\usepackage{fancybox}
\usepackage{color}
\usepackage{ulem,bbold}
\usepackage{enumitem}
\usepackage{subfigure}
\usepackage{bbm}
\usepackage{parskip}
\usepackage{cite}
\usepackage{cite}
\newcommand{\Comment}[1]{{}}
\definecolor{MyDarkBlue}{rgb}{0.15,0.15,0.45}
\usepackage[linktocpage=true]{hyperref}
\hypersetup{
colorlinks=true,
citecolor=MyDarkBlue,
linkcolor=MyDarkBlue,
urlcolor=MyDarkBlue,
}

\newcommand{\be}{\begin{equation}}
\newcommand{\ee}{\end{equation}}
\newcommand{\bea}{\begin{eqnarray}}
\newcommand{\eea}{\end{eqnarray}}
\newcommand{\beas}{\begin{eqnarray*}}
\newcommand{\eeas}{\end{eqnarray*}}
\newcommand{\nn}{\nonumber}

\newcommand{\llp}{\left [}
\newcommand{\rrp}{\right ]}
\newcommand{\lp}{\left (}
\newcommand{\rp}{\right )}

\newcommand{\dd}{\text{d}}
\newcommand{\W}{\mathcal{W}}
\newcommand{\TD}{\tilde{\nabla}}

\newcommand{\rz}{\rho_0}

\numberwithin{equation}{section}

\begin{document}


\begin{center}
{\Large \bf{Multi-Galileons in Curved Space\\}}
\vspace{1cm}
\end{center}

\vspace{1truecm}
\thispagestyle{empty}
\centerline{\Large Alice Garoffolo,${}^{\rm a,}$\footnote{\href{mailto:aligaro@sas.upenn.edu}{\texttt{aligaro@sas.upenn.edu}}} Kurt Hinterbichler,${}^{\rm b,}$\footnote{\href{mailto:kurt.hinterbichler@case.edu} {\texttt{kurt.hinterbichler@case.edu}}} Mark Trodden,${}^{\rm a,}$\footnote{\href{mailto:trodden@upenn.edu} {\texttt{trodden@upenn.edu}}} }

\vspace{.5cm}

\centerline{{\it ${}^{\rm a}$Center for Particle Cosmology, Department of Physics and Astronomy,}}
\centerline{{\it University of Pennsylvania, Philadelphia, Pennsylvania 19104, USA }}
\vspace{.25cm}

\centerline{{\it ${}^{\rm b}$CERCA, Department of Physics,}}
\centerline{{\it Case Western Reserve University, 10900 Euclid Ave, Cleveland, OH 44106}} 
\vspace{.25cm}

\begin{abstract} 
Using the probe brane construction of higher derivative effective field theories, extended to higher co-dimensions and curved spaces, we construct galileon and DBI theories on de Sitter space with $N$ fields and an $\frak{so}(N)$ internal symmetry, non-linearly realizing the symmetries of a higher dimensional de Sitter space.  In some cases, the theory admits a non-trivial vacuum that spontaneously breaks the $\frak{so}(N)$ symmetry, and around this vacuum the Goldstone modes have vanishing kinetic terms and become infinitely strongly coupled.  This gives an example of a scalar effective field theory with two de Sitter vacua, one of which appears to have Boulware-Deser-like ghosts, and one of which does not.

\end{abstract}

\newpage

\thispagestyle{empty}
\tableofcontents

\setcounter{page}{1}
\setcounter{footnote}{0}

\parskip=5pt
\normalsize

\section{Introduction}

The probe brane construction is a convenient framework for generating a diverse landscape of scalar effective field theories (EFTs) by embedding lower-dimensional branes into fixed higher-dimensional bulk spacetimes. In this approach the scalar fields have a geometric interpretation as fluctuations, or brane bending modes, of the brane within the higher-dimensional space.   If $D$ and $d$ are the dimensions of the bulk and brane respectively, then the number of independent scalar fields that emerge in the resulting theory on the brane is given by the co-dimension $N=D-d$.  Any isometry of the bulk space will be realized as global symmetries of the resulting theory.  Those that preserve the ground state configuration of the brane will be unbroken, and those that move the brane will be spontaneously broken.
The power of this construction lies in its versatility: by judiciously selecting the dimensions and the geometries of both the bulk and the brane, a wide spectrum of scalar field theories with various spontaneously broken and unbroken symmetries can be realized.

The probe brane construction can reproduce many known and well-studied EFTs that were originally found from other considerations, including the Dirac-Born-Infeld (DBI) galileons and the original galileons \cite{Nicolis:2008in} that appear in its small field limit \cite{deRham:2010eu}, the special galileon \cite{Novotny:2016jkh} (originally studied in \cite{Cheung:2014dqa,Cachazo:2014xea,Hinterbichler:2015pqa}), and gauged galileons \cite{Goon:2012mu} (originally studied in \cite{Zhou:2011ix}).  In addition, it has been used to find new theories, such as the galileons on de Sitter (dS) space~\cite{Goon:2011qf,Goon:2011xf,Goon:2011uw,Burrage:2011bt} (which are examples of shift symmetric theories \cite{Bonifacio:2018zex,Bonifacio:2021mrf} realizing the discrete series of unitary dS representations), galileons on more general cosmological spaces \cite{Goon:2011xf}, and multi-field galileons ~\cite{Hinterbichler:2010xn,Yu:2024sed} on flat space (which were also found and studied by other methods in \cite{Deffayet:2010zh,Padilla:2010de,Padilla:2010tj,Padilla:2010ir,Zhou:2010di,Goon:2012dy,Allys:2016hfl,Klein:2017npd,Bogers:2018kuw,Roest:2019oiw,Kampf:2020tne,Aoki:2021kla}).  See \cite{Trodden:2011xh,Deffayet:2013lga} for earlier reviews of some of these generalizations.

A crucial property of the above mentioned theories is that, despite the presence of higher-order derivatives in their Lagrangians, their equations of motion remain of second order, avoiding the non-linear propagation of extra, generally ghostly, degrees of freedom. In the probe brane construction, the origin of this property is directly traced to the fact that the galileon action is derived from Lovelock invariants of the induced metric on the brane and boundary terms associated with the bulk Lovelock invariants~\cite{deRham:2010eu}.

The goal of this paper is to explore the probe brane construction in situations where both the brane and bulk are curved, and in which the co-dimension $N$ is arbitrary.  This extends the setup of \cite{Hinterbichler:2010xn} to curved space, and that of \cite{Goon:2011qf,Goon:2011xf} to higher co-dimension, and will lead to new galileon-like theories on curved space containing multiple scalar fields that non-linearly realize larger symmetry groups.  

Our primary example will be a theory constructed from embedding a dS$_d$ brane into a higher dimensional dS$_D$, with an $\frak{so}(N)$ rotational symmetry about the co-dimension $N=D-d$ brane.  This gives rise to a theory with $N$ scalars living on dS$_d$.  The full symmetry algebra is the algebra of isometries of dS$_D$, which is $\frak{so}(1,D)$.   Of these, the isometries of the dS$_d$ brane form an $\frak{so}(1,d)$ subalgebra that is linearly realized.  The $\frak{so}(N)$ rotational symmetry in the extra dimensions forms an $\frak{so}(N)$ subalgebra that is also linearly realized, with the $N$ scalars transforming in the fundamental.  The remaining generators are all non-linearly realized as extended shift symmetries for the scalars.  At linear order they look like $N$ independent dS galileons, with the mass fixed by the galileon shift symmetries ($k=1$ shift symmetries in the language of \cite{Bonifacio:2018zex,Bonifacio:2021mrf}).

The multi-galileon action is more constrained than its single-field counterpart.  In $d=4$ there are (at least for even co-dimension) only two terms in the relevant galileon Lagrangian, as opposed to five in the single field case.  In our example, these two terms give rise to an $\frak{so}(N)$ invariant potential for the scalars whose shape depends on the parameters in front of the terms.  For some choices of these parameters, there can be additional dS invariant vacua that develop away from the origin in field space, spontaneously breaking the internal $\frak{so}(N)$ symmetry to $\frak{so}(N-1)$.  Such breaking of an internal symmetry should give rise to $N-1$ Goldstone bosons.  We will see that the Goldstones are present in cubic and higher order interactions, but have vanishing kinetic terms, and so they are infinitely strongly coupled.

A theory that propagates more degrees of freedom non-linearly than it does at linear order around some vacuum solution is often said to have a Boulware-Deser ghost, after \cite{Boulware:1972yco} which illustrated this phenomenon in generic interacting theories of massive gravity.  In the case of massive gravity, much effort went into constructing special interactions that evade this issue \cite{deRham:2010kj,Hassan:2011hr} (see \cite{Hinterbichler:2011tt,deRham:2014zqa} for reviews), so that the non-linear theory has the same number of degrees of freedom as the linear theory.  In our case, looking around an $\frak{so}(N)$ breaking vacuum, there is only one propagating mode, whereas the full non-linear theory has $N$ degrees of freedom, so there are Boulware-Deser-like ghosts with respect to this vacuum.  However, in this same theory, there is another dS vacuum, the $\frak{so}(N)$ invariant one, where all $N$ degrees of freedom propagate in a normal way.  This gives an explicit and fully dS invariant illustration of another way in which Boulware-Deser-type ghosts can be rendered harmless: not by reducing the number of non-linear degrees of freedom, but by finding another vacuum in which all the degrees of freedom propagate as healthy modes.

{\bf Conventions:} We use the metric signature $(-,+,+,\ldots)$, and the curvature conventions are those of \cite{Carroll:2004st}.

\section{Probe brane construction of the multi-field action\label{PBconstructionsection}}

In this section, we give a brief overview of the probe brane construction, introducing the geometric quantities that we will need.  For more details, especially regarding the generalization to higher co-dimensions, we refer the reader to the Appendix of~\cite{Hinterbichler:2010xn}.

We consider the theory of a brane of spacetime dimension $d$ probing a fixed bulk background of spacetime dimension $D$.
The co-dimension is the difference $N = D - d$, and it will give the number of dynamical scalar fields in the theory.
Capital Latin letters $\{A, B, \dots \}$ are used as indices for the bulk coordinates $X^A$ and run from $0$ to $D-1$, while 
Greek letters $\{\mu, \nu, \dots  \}$ are used for the brane coordinates $x^\mu$ and run from $0$ to $d-1$.  There is a fixed non-dynamical bulk metric given by $G_{AB}(X)$.

The brane is embedded via the functions $X^A (x)$.  There are $d$ tangent vectors and $N$ normal vectors to the brane, $\{ e^A_\mu, n^A_i \}$ respectively.  Here $\{i,j,\ldots\}$ run over the number of co-dimensions and are associated with an orthonormal basis in the normal bundle of the brane. 
The tangent vectors are defined through 
\be \label{eq:TangentVectors}
e^A_\mu \equiv \frac{\partial X^A}{\partial x^\mu}\,,
\ee 
while the normal vectors can be chosen to be any set compatible with the orthonormality relations,
\be 
G_{A B} \, n^A_{\ i} n^B_{\, j} = \delta_{ij} \,, \qquad G_{AB} \,  e^A_{\ \mu} n^B_{\ i} = 0\,.
\ee 

From these we can construct the various geometric quantities defined on the brane. The induced metric and extrinsic curvatures are
\be 
\bar g_{\mu\nu} = e^A_\mu e^B_\nu G_{AB} \,, \qquad K^i_{\mu\nu} =  e^A_\mu e^B_\nu \nabla_B n_A^i\, , \label{inducefeefe}
\ee 
where indices are raised and lowered by $G_{AB}$ and $\eta_{ij}$ and their inverses, so that $  n^A_{\ i} = \eta_{ij} G^{AB} n_B^{\  j}$.
The induced metric $\bar g_{\mu\nu}$ can be used to build a connection on the tangent bundle as usual, via
\be
   \bar \Gamma^\rho_{\mu\nu}  \equiv e^{\ \rho}_A e^B_{\ \mu} \nabla_B e^A_{\ \nu} = \frac{1}{2} \bar g^{\rho\sigma} \lp \partial_\mu \bar g_{\sigma \nu} + \partial_\nu \bar g_{\sigma \mu} - \partial_\sigma \bar g_{\mu\nu}\rp\,,
\ee 
from which the action of the covariant derivative $\bar\nabla_\mu$ on tangent indices $\{\mu,\nu,\ldots\}$, and the brane curvature 
\be
    \bar R^\rho_{\ \sigma \mu \nu} = \partial_\mu \bar\Gamma^\rho_{\sigma\nu} - \partial_\nu\bar \Gamma^\rho_{\sigma\mu} +\bar \Gamma^\rho_{\mu\lambda} \bar\Gamma^\lambda_{\sigma \nu} - \bar\Gamma^{\rho}_{\nu \lambda} \bar \Gamma^{\lambda}_{\sigma\mu}\,,
\ee 
are defined.
Additionally, for co-dimension $N\geq 2$ there can be a non-trivial { twist connection}, defined by
\be
\beta^i_{\ j \mu} \equiv n^i_{\ A} e^{B}_{\ \mu} \nabla_B n^A_{\ j}\,.
\ee
This is the metric connection on the normal bundle to the brane.  The twist connection enters when the covariant derivative $\bar\nabla_\mu$ acts on normal bundle indices $\{i,j,\ldots\}$, and defines a curvature in the normal bundle,
\be
{}^{(\perp)}R^{i}_{\ j \mu \nu} = \partial_\mu \beta^i_{\ j \nu} - \partial_\nu \beta^i_{\ j \mu} + \beta^{i}_{\ \mu k} \beta^{k}_{\ \nu j} -\beta^{i}_{\ \nu k} \beta^{k}_{\ \mu j}\,.
\ee 
These geometric ingredients are the building blocks of the action for the scalar fields representing the bending modes of the brane floating in the bulk.   The most general brane Lagrangian is a diffeomorphism scalar constructed from these ingredients, with the embedding functions $X^A$ as the dynamical variables:
\be\label{eq:ActionGeneric}
{ S} = \int d^d x \sqrt{-  \bar g} \, {\cal L} \lp  \bar g_{\mu\nu}, \bar\nabla_\mu,  \bar R^\rho_{\ \sigma\mu\nu}, {}^{(\perp)} R^i_{\ j\mu\nu}, K^i_{\mu\nu}\rp\,. 
\ee 
Note that the geometric ingredients are not always independent, as there can be Gauss-Codazzi type equations that relate them through the bulk curvature (see e.g. footnote \ref{footnote1label}).

\subsection{Gauge fixing and symmetries}

Since the action \eqref{eq:ActionGeneric} is a diffeomorphism scalar, it is invariant under reparametrizations of the brane coordinates $x^\mu$.  This is a gauge symmetry, infinitesimally given by
\be\label{eq:GaugeDiffeo}
\delta_g X^A = \xi^\mu \partial_\mu X^A\,,
\ee 
where $\xi^\mu(x)$ is the gauge parameter.

The action \eqref{eq:ActionGeneric} inherits global symmetries from any Killing symmetry of the higher-dimensional bulk.  If $K^A(X)$ is a Killing vector of the bulk metric,
\be\label{eq:KillingEq}
K^C \partial_C G_{AB} + G_{CB} \partial_A K^C   + G_{CA} \partial_B K^C = 0\,,
\ee 
then the induced metric, the extrinsic curvatures, and thus all the the geometric quantities, are invariant under the transformation
\be\label{eq:KillingTransf}
\delta_K X^{A} = K^A (X)\,.
\ee
These are therefore global symmetries of the action \eqref{eq:ActionGeneric}.

We may fully fix the gauge freedom \eqref{eq:GaugeDiffeo} by breaking up the bulk directions $A\rightarrow (\mu,I)$ with $\mu\in \{ 0,1,\ldots,d-1\}$, $I\in \{1,\ldots,N\}$ and demanding
\be\label{eq:GaugeFixingDiffeo}
X^A (x) = \left\{
\begin{matrix}
    x^\mu \qquad & \mbox{if} \qquad A = \mu\, , \\
    \pi^I(x) \qquad &\mbox{if} \qquad A = I \, , \\
\end{matrix} \right. \,
\ee 
so that the world-volume coordinates of the brane are fixed to the first $d$ bulk coordinates, which can be arbitrary.
In this gauge, hypersurfaces at $X^I =\pi^I= {const.}$ foliate the bulk,  and varying $\pi^I$ fields describe the brane bending modes of the brane embedded in the bulk.

Now, the gauge choice \eqref{eq:GaugeFixingDiffeo} is not generally preserved by a global transformation \eqref{eq:KillingTransf}. However, a compensating gauge transformation can be added to restore the gauge choice.
Consequently, the gauged fixed action will be invariant under a combination of two transformations:
one generated by a Killing vector as in Eq.~\eqref{eq:KillingTransf}, and the other a gauge transformation as in Eq.~\eqref{eq:GaugeDiffeo}, with the generator chosen as $\xi^\mu = - K^\mu$, which restores the gauge choice \eqref{eq:GaugeFixingDiffeo} (indeed, under the sum of these two transformations we have $
\lp \delta_K + \delta_\xi \rp X^\mu  = K^\mu - K^\nu \partial_\nu x^\mu = K^\mu - K^\mu = 0$).
The $\pi^I$ fields, under the sum of these transformations, transform according to $
\lp \delta_K + \delta_\xi \rp X^I =  K^I(x, \pi) - K^\nu (x, \pi) \, \partial_\nu X^I = K^I(x, \pi) - K^\nu(x, \pi) \partial_\nu \pi^I\,$.
Hence the gauged fixed action of the scalar field multiplet will be invariant under the transformation:
\be \label{eq:PiITransformation}
\delta \pi^I = K^I(x, \pi) - K^\mu(x, \pi) \partial_\mu \pi^I\,.
\ee

\subsection{Galileon terms}

In addition to symmetry considerations, we also wish to build a theory with second order equations of motion.
This is achieved by choosing Lovelock invariants \cite{Lovelock:1971yv} as building blocks of the action \cite{deRham:2010eu}.  These include Lovelock invariants made from the induced metric $\bar g_{\mu\nu}$ in \eqref{inducefeefe}, as well as boundary terms coming from bulk Lovelock invariants~\cite{Myers:1987yn,Miskovic:2007mg}, generalizing the Gibbons-Hawking term for Einstein gravity \cite{Gibbons:1976ue,York:1972sj}.  For $N=1$, there are $d+1$ such invariants, which give rise to DBI galileons \cite{deRham:2010eu}.  For $N\geq 2$, their classification is more involved \cite{Charmousis:2005ey,Charmousis:2005ez} (see the discussion in \cite{Hinterbichler:2010xn}).  The end result is that in $d=4$, {and at least in the case of even $N$}, there are only 2 terms that are independent, the brane cosmological constant and brane Einstein-Hilbert term\footnote{In some case there is a boundary term $\sim K^i K_i + K^i_{\ \mu\nu}K_i^{\ \mu \nu} $ known as the Myers boundary term ~\cite{Myers:1987yn,Miskovic:2007mg}.  In the case of maximally symmetric bulks, which is the case we will ultimately be interested in, this is not independent because it can be related directly to the intrinsic curvature using the Gauss-Codazzi equation generalized to higher co-dimensions,
\be\label{eq:GaussCodazzi}
R^{(D)}_{ABCD}e^{A}_{\ \rho}e^{B}_{\ \sigma} e^{C}_{\ \mu}e^{D}_{\ \nu} = \bar R^{(d)}_{\rho\sigma \mu\nu} + K^i_{\ \mu\sigma}K_{i \nu \rho} - K^i_{\ \nu\sigma}K_{i \mu \rho}\,,
\ee 
where $R^{(D)}_{ABCD}$ is the Riemann tensor built out of $G_{AB}$, $\bar R_{\rho\sigma \mu\nu}$ is the Riemann tensor built out of $\bar g_{\mu\nu}$.

If the bulk is a maximally symmetric spacetime, its Riemann tensor is determined in terms of the metric by
\be\label{eq:RiemannMaximallySymmetric}
    R^{(D)}_{ABCD} = \frac{R^{(D)}}{D(D-1)} \lp G_{AC} G_{BD} - G_{BC} G_{AD}\rp \,,
\ee 
with $R^{(D)}$ the constant bulk Ricci scalar,
allowing one to compute the contractions between the Riemann tensor and the tangent vectors $e^A_{\ \mu}$ appearing on the left hand side of \eqref{eq:GaussCodazzi},
\begin{align}\label{eq:RiemannDMaximallySymmetric}
    R^{(D)}_{ABCD} e^{A}_{\ \rho}e^{B}_{\ \sigma} e^{C}_{\ \mu}e^{D}_{\ \nu} 
    &= \frac{R^{(D)}}{D(D-1)} \lp \bar g_{\rho \mu} \bar g_{\sigma \nu} - \bar g_{\rho \nu} \bar g_{\sigma \mu} \rp \,.
\end{align}
One can then find a relation between the Myers term and the Einstein-Hilbert term by taking the contraction of Eq.~\eqref{eq:GaussCodazzi} with $\bar g^{\rho \mu} \bar g^{\sigma \nu}$,
\begin{eqnarray}
   \bar R - K^i K_i + K^i_{\ \mu\nu}K_i^{\ \mu \nu}  &=& R^{(D)} \frac{d(d-1)}{D(D-1)} \,.
\end{eqnarray}
We see that the Myers term is the same as the Einstein-Hilbert up to an additive constant and hence brings no additional term to the action.  \label{footnote1label}
}.  
Restricting to this case, we will thus take our action to be
\be S= \int d^d x \sqrt{-\bar g} \lp -a_2 + a_4 \bar R \rp\,, \label{fullactioneqe}\ee
where $a_2$, $a_4$ are free coefficients of mass dimension $d$ and $d-2$ respectively.  We will keep $d$ general for the most part, fixing $d=4$ when we get to the explicit example in section~\ref{sec:ExplicitExample}.

\subsection{Warped product metric\label{wpmsection}}

We will assume that the bulk metric is in warped product form
\be \label{eq:MetricWarped}
G_{AB} \, \dd X^A \dd X^B =  e^{2 \W(\omega)}\, {g}_{\mu\nu}(x) \dd x^\mu \dd x^\nu + H_{IJ}(\omega) \dd \omega^I \dd \omega^J\,, 
\ee 
where we have separated the bulk coordinate into brane and normal coordinates as $X^A = \{ x^\mu, \omega^I \}$. 
Here ${g}_{\mu\nu}(x)$ is a $d$ dimensional metric depending only on the $x^\mu$ coordinates, while the warp factor $\W(\omega)$, and  the metric $H_{IJ}(\omega)$ in the normal directions, depend only on the $\omega^I$ coordinates. 
Note that this form of the metric is more general compared to the ones considered in~\cite{Goon:2011qf,Goon:2011xf,deRham:2010eu,Hinterbichler:2010xn,Burrage:2011bt}.  The cases in ~\cite{Goon:2011qf,Goon:2011xf,Burrage:2011bt} correspond to $N=1$, and those in \cite{Hinterbichler:2010xn} correspond to the special case where $H_{IJ}=\delta_{IJ}$ and $\W=0$.  

In the gauge choice \eqref{eq:GaugeFixingDiffeo}, the induced metric \eqref{inducefeefe} reads
\be
 \bar g_{\mu\nu} = e^{2 \W(\pi)}  g_{\mu\nu}(x) + H_{IJ} (\pi)  \partial_\mu \pi^I \partial_\nu \pi^J\,.\label{genwrpedmetrhree}
\ee

The global symmetry transformations \eqref{eq:PiITransformation} are linearly or non-linearly realized depending on whether the relevant Killing vectors preserve the brane or not.  Consider a Killing vector $K^A_{\|}$ which is tangent to the brane, so that $K^I_{\|}=0$.  The $I\mu$ components of the Killing equation \eqref{eq:KillingEq} tell us that $\partial_I K^\mu_{\|}=0$, so that $K^\mu_{\|}$ is independent of $\pi^I$.    The $\mu\nu$ components of the Killing equation then tell us that $K^\mu_{\|}$ is a Killing vector of the brane metric $g_{\mu\nu}$.  (The $IJ$ Killing equation is identically satisfied and gives no further constraint.)   
The transformation \eqref{eq:PiITransformation} then becomes
\be \label{eq:PiITransformation2}
\delta \pi^I = - K^\mu_{\|}(x) \partial_\mu \pi^I\,,
\ee
which shows that the symmetry is linearly realized, with the $\pi^I$ transforming as scalars under the $d$-dimensional isometries generated by $K^\mu_{\|}$.

Next, consider a Killing vector $K^A_{\perp}$ which is perpendicular to the brane, so that $K^\mu_{\perp}=0$. The $I\mu$ components of the Killing equation \eqref{eq:KillingEq} tell us that $\partial_\mu K^I_{\perp}=0$, so that $K^I_{\perp}$ is independent of $x^\mu$.    The $\mu\nu$ Killing equation then implies that $K^I_{\perp}\partial_I\W=0$, so that the warp factor must be constant along the direction of the Killing vector.   The $IJ$ components of the Killing equation \eqref{eq:KillingEq} then tell us that $K^I_{\perp}$ is a Killing vector of the metric $H_{IJ}$. 
The transformation \eqref{eq:PiITransformation} thus becomes
\be \label{eq:PiITransformation3}
\delta \pi^I = K^I_\perp( \pi) \,.
\ee
This is the transformation law of a sigma model with target space parametrized by $\pi^I$ and target space metric $H_{IJ}$.  Those $K^I_\perp( \pi)$ with an expansion in $\pi^I$ that starts at linear order will be linearly realized, the rest will be non-linearly realized.  

Finally, a generic Killing vector which is neither tangent nor transverse will give a transformation which is non-linearly realized.

\subsection{Action and derivative expansion}\label{sec:DerivativeExpansion}

To derive the action of the $\pi^I$ fields, we substitute the expression \eqref{genwrpedmetrhree} for the induced metric into the action \eqref{fullactioneqe}.  
When $a_4=0$ we can write the action in closed form, giving the multi-field DBI theory for a brane probing a warped geometry,  
\be   
{\cal L}= -a_2  \sqrt{-\det \left( e^{2\W} g_{\mu\nu}+ H_{IJ}  \nabla_\mu \pi^I\nabla_\nu \pi^J\right)}\,,
\ee
which has been extensively studied in the context of stringy models of the inflationary early universe \cite{Huang:2007hh,Langlois:2008qf,Langlois:2008wt,Langlois:2009ej,Mizuno:2009mv}.

When $a_4\not=0$ the action in closed form can be computed for a given $d$ (see appendix A of \cite{Renaux-Petel:2011rmu} for the $d=4$ case, using results from \cite{silva2000traceformulasyieldinverse}), but the result is lengthy and we will not need it.  We will be able to see everything we need by working in an expansion in powers of derivatives.  In our example below, to find the quartic galileon, which is the first non-trivial galileon beyond the kinetic term, we will need the derivative expansion up to sixth order in derivatives.  The resulting expressions are lengthy so we relegate them to appendix \ref{sec:CC}.

Up to second order in powers of derivatives, and ignoring total derivatives, the Lagrangian is
\be 
{ 1\over \sqrt{-g}} {\cal L}  = - e^{d\W} a_2 + e^{(d-2)\W}\left[  a_4 R- {1\over 2}\left(  a_2 H_{IJ} -2 a_4 (d-1)(d-2) \partial_I\W\partial_J\W\right) \nabla_\mu \pi^I\nabla^\mu \pi^J \right]+{\cal O}\left(\partial^4\right)\,,\label{expabdseciree}
\ee
where $R$ is the Ricci scalar of the brane metric $g_{\mu\nu}$.
DBI models including a brane Einstein-Hilbert term have also been of interest for inflationary cosmology, partly because of their unusual features such as the ability to generate large non-gaussianities of orthogonal shape \cite{Renaux-Petel:2011rmu,Renaux-Petel:2011lur,Fasiello:2013dla}.

The expansion to second order shown in \eqref{expabdseciree} is all that is needed to see the scalar potential in a fixed background, since all higher powers of the derivatives vanish when $\pi^I=const.$  The potential reads
\be 
V(\pi)=-\left.{\cal L}\right|_{\pi^I=const.}=  a_2  e^{d\W} - a_4 e^{(d-2)\W} R \,. \label{genpotentiale}
\ee
The potential depends on the fields through warp factor $\W(\pi)$, and also depends on the background curvature $R$.  In particular, if the warp factor is non-trivial, the scalar fields can acquire a mass depending on the shape of the warp factor, including possible contributions from the brane curvature if $d\neq2$.

\section{de Sitter example}\label{sec:ExplicitExample}

In the previous section, we have discussed the general form for the action of the multiple scalar fields, $\pi^I$, including an Einstein-Hilbert term on the brane.  Its derivative expansion up to six derivative terms are given in Eqs.~\eqref{eq:CCintermsofPI} and \eqref{eq:EHintermsofPI}. 

We now turn to a specific example, one which maximizes the amount of symmetry.   We consider the bulk and the brane to be de Sitter spacetimes of dimensions $D$ and $d$ respectively.  In addition, we want to maintain $\frak{so}(N)$ rotational symmetry about the brane in the extra $N=D-d$ dimensions, which will show up as an internal flavor symmetry in the scalar theory.
When presenting explicit results for the action, we will fix $d=4$.

\subsection{Bulk metric in warped form}

The first task is to write the bulk dS$_D$ metric as a warped product of a brane dS$_d$ metric and some $\frak{so}(N)$ invariant metric in the orthogonal directions, of the form \eqref{eq:MetricWarped}.
To achieve this, we embed the bulk itself into an auxiliary ($D+1$)-dimensional flat Minkowski spacetime with metric $\eta_{{\cal A} {\cal B}} = {\rm diag} \lp \eta_{AB} , \delta_{IJ} \rp$, where $A\in \{0,\dots,d\}$, $I \in \{1,\dots,N\}$ and $\eta_{AB}$ is a flat $(d+1)$ dimensional Minkowski metric. 
Here the calligraphic indices $\{{\cal A},{\cal B},\ldots\}$ run over the coordinates $ Y^{\cal{A}}$ in the $(D+1)$-dimensional ambient space.

Let $X^A(x)$ be an embedding of $d-$dimensional de Sitter $({\rm dS}_d)$ spacetime into a flat Minkowski $d+1$ space, such that $\eta_{AB} X^A X^B = L^2_d$, and let $n^I(\Omega)$ be an embedding of the unit $N-1$ sphere into euclidean $N$ space such that $\delta_{IJ} n^I n^J=1$. 
Now combine these into the embedding 
\be
   Y^{{\cal A}}  =\begin{cases} F_d (\rho) \, X^A(x)  &   {\rm if} \qquad {\cal A} = A\,, \\
   L_d \, F_N (\rho) \, n^I(\Omega) & {\rm if} \qquad {\cal A} = I \,.
   \end{cases}   \label{eq:embeddingDSDplus1Pis}
\ee
Here $\rho$ is a dimensionless radial coordinate in the extra dimensions, and $F_d (\rho)$, $F_N (\rho)$ are dimensionless functions of this radial coordinate that we will soon determine.  Requiring  that the full $D= d + N $ dimensional space is ${\rm dS}_D$ with radius $L_D$, we impose $\eta_{{\cal A}{\cal B}}Y^{\cal A} Y^{\cal B}={1\over L_D^2}$, which determines a relation between $F_N$ and $F_d$, 
\be\label{eq:FNintermsofFd}
L^2_d \lp F^2_d + F^2_N\rp = L^2_D\,.
\ee  
The induced metric computed from the embedding \eqref{eq:embeddingDSDplus1Pis} is 
\be
\dd s^2 = F^2_d \,  g_{\mu\nu} (x)\dd x^\mu \dd x^\nu + L^2_d \lp {F'_d}^2+{F'_N}^2\rp \dd \rho^2 + L^2_d F^2_N \dd \Omega^2_{N-1}\,,\label{indcafregede}
\ee 
with primes denoting derivatives with respect to $\rho$. Here the $d-$dimensional metric $g_{\mu\nu}$ is a ${\rm dS}_d$ metric with radius $L^2_d$. 

There is a redundancy due to reparameterizations of the radial coordinate $\rho$, which we will fix by demanding that the $N$ dimensional part of the metric \eqref{indcafregede} be manifestly conformally flat. Thus we require
\be
L^2_d F^2_N = \rho^2 L^2_d \lp {F'_d}^2+{F'_N}^2\rp\,.
\ee 
After using \eqref{eq:FNintermsofFd} to eliminate $F_N(\rho)$, this gives a first order differential equation for $F_d(\rho)$,
\be
F^2_d + \frac{\rho^2 {F'_d}^2}{1 - \frac{L^2_d}{L^2_D} F^2_d} = \frac{L^2_D}{L^2_d}\,.
\ee 
The solution that is positive at small $\rho$ is (the most general such solution has a free parameter, but we can absorb it by rescaling $\rho$),
\be
F_d(\rho) = \frac{L_D}{L_d} \, \frac{1-\rho^2}{1+\rho^2}\,.\label{Fdefine1}
\ee 
Solving \eqref{eq:FNintermsofFd} for a positive $F_N(\rho)$ then gives
\be F_N(\rho)={L_D\over L_d} {2\rho\over 1+\rho^2}\,, \label{Fdefine2}
\ee
and thus the explicit form of the bulk metric is fixed to be
\be\label{eq:MetricExample}
\dd s^2 = G_{A B} \dd X^A \dd X^B =  L^2_D \llp \, \lp \frac{1 - \rho^2 }{1 +  \rho^2}\rp^2 \frac{1}{L^2_d} \,  g_{\mu\nu} \dd x^\mu \dd x^\nu + \frac{4}{(1+\rho^2)^2} \lp \dd \rho^2 + \rho^2 \dd \Omega^2_{N-1} \rp  \rrp\,.
\ee 
The metric is regular at $\rho=0$ and approaches a coordinate horizon at $\rho=1$; we will confine ourselves to the region $0\leq \rho<1$.

This metric describes a $d$ dimensional dS space embedded into a $(d+N)$-dimensional dS space, in such a way that there is full $\frak{so}(N)$ rotational symmetry among the extra dimensions.  From the general discussion in section \ref{wpmsection}, we can expect that the symmetries acting on the $\pi^I$ will realize a symmetry breaking pattern $\frak{so}(1,D) \rightarrow  \frak{so}(1,d)\oplus  \frak{so}(N)$.

\subsection{Killing vectors and symmetries}

As discussed previously, the bulk Killing vectors give the symmetry transformations of the theory.  The dS$_D$ metric in Eq.~\eqref{eq:MetricExample} is maximally symmetric, guaranteeing the maximum number of symmetries in the resulting scalar field theory. 
We can find the Killing vectors and the resulting symmetries by exploiting the bulk embedding \eqref{eq:embeddingDSDplus1Pis} into a $D+1$ dimensional flat spacetime. 

To do this explicitly, we use conformal inflationary coordinates $x^\mu=(\tau,y^i)$, $i \in\{ 1 , \dots,  d-1$\}, on the de Sitter brane, so that $g_{\mu\nu}dx^\mu dx^\nu=\frac{1}{L_d^2\tau^2} \lp - \dd \tau^2 + \delta_{ij} \dd y^i \dd y^j \rp$, and we change to cartesian coordinates $\omega^I$, $I \in \{ 1, \dots, N\}$, in the orthogonal directions, so that the bulk metric is
\be
\dd s^2 = G_{A B} \dd X^A \dd X^B =  L^2_D \llp \, \lp \frac{1 - \rho^2 }{1 +  \rho^2}\rp^2  \frac{1}{\tau^2} \lp - \dd \tau^2 + \delta_{ij} \dd y^i \dd y^j \rp+ \frac{4 \, \delta_{IJ} \dd \omega^I \dd \omega^J}{(1+\rho^2)^2} \rrp\,,
\ee 
where $ \rho = \sqrt{ \delta_{IJ} \omega^I \omega^J}$.
In these coordinates, the embeddings in \eqref{eq:embeddingDSDplus1Pis} are
\begin{align}
    Y^{0} &=  L_d\, F_d(\rho) \, \frac{1 - \tau^2 + y^2 }{2 \tau} \,,  \nn \\
    Y^{1} &= L_d \, F_d(\rho) \, \frac{1 + \tau^2 - y^2 }{2 \tau}\,, \nn\\
    Y^{i} &=  L_d\, F_d(\rho) \, \frac{y^i}{\tau}\, ,\nn \\
    Y^{I} &= L_d \,  F_N(\rho) \, \frac{\omega^I}{\rho}  \,, \label{expebfsfee}
\end{align}
where $y^2 \equiv \delta_{ij} y^i y^j$ and $F_d$, $F_N$ are as in \eqref{Fdefine1}, \eqref{Fdefine2}.

The bulk Killing vectors can be obtained by pulling back those Killing vectors of the ambient $(D+1)$-dimensional Minkowski space ${\rm M}_{D+1}$ that preserve the ${\rm dS}_D$ hypersuface.
These are the $D(D+1)/2$ generators corresponding to the Lorentz transformations of ${\rm M}_{D+1}$,
\be
{\cal M}_{{\cal A} {\cal B}}  = \eta_{{\cal A}{\cal C}} Y^{{\cal C}} \frac{\partial}{\partial Y^{{\cal B}}}  -  \eta_{{\cal B}{\cal C}} Y^{\cal C} \frac{\partial}{\partial Y^{{\cal A}}}\equiv {\cal M}^{\cal C}_{{\cal A}{\cal B}} \frac{\partial }{\partial Y^{\cal C}}  \,.
\ee 
The component expressions ${\cal M}^A_{{\cal A}{\cal B}}$ of the Killing vectors in the intrinsic coordinates $X^A$ of dS$_D$ can be obtained by pulling back using the embedding \eqref{expebfsfee},
\be
{\cal M}^A_{{\cal A}{\cal B}} = G^{AB} \, \frac{\partial Y^{\cal C}}{\partial X^B} {\cal M}^{\cal D}_{{\cal A}{\cal B}} \eta_{{\cal C}{\cal D}}  \,.
\ee 

Consider first the Killing vectors
\begin{align}
    {\cal M}_{10} =      Y^{1} \frac{\partial}{\partial Y^{0}} + Y^{0} \frac{\partial}{\partial Y^{1}} \quad &\to \quad d \equiv -\tau \, \partial_\tau- y^i \, \partial_i \,, \nn\\ 
    {\cal M}_{0i} = - Y^{0} \frac{\partial}{\partial Y^{i}} -   Y_{i} \frac{\partial}{\partial Y^{0}}  \quad &\to \quad j^{(0)}_i \equiv \tau y_i  \partial_\tau + \frac12 \lp  \tau^2 - y^2 -1\rp \partial_i + y_i y^j \partial_j \, , \nn\\ 
    {\cal M}_{1j} = Y^{1} \frac{\partial}{\partial Y_{j}} -   Y_{j} \frac{\partial}{\partial Y^{1}} \quad &\to \quad j^{(1)}_i \equiv \tau y_i  \partial_\tau + \frac12 \lp  \tau^2 - y^2 +1 \rp \partial_i + y_i y^j \partial_j \, , \nn\\ 
    {\cal M}_{ij} = Y_{i} \frac{\partial}{\partial Y^{j}} -   Y_{j} \frac{\partial}{\partial Y^{i}} \quad &\to \quad j_{ij} \equiv y_i \partial_j - y_j \partial_i\,.
\end{align}
These are the Killing vectors tangent to the brane and, as such, only depend on the brane coordinates.  They generate the linearly realized dS$_d$ transformations of the theory, according to \eqref{eq:PiITransformation2}.   Taking the linear combinations 
\be 
p_i= j^{(0)}_i - j^{(1)}_i  =-\partial_i,\ \ \ k_i=-\left( j^{(1)}_i + j^{(0)}_i \right)=-2\tau y_i  \partial_\tau -\lp  \tau^2 - y^2 \rp \partial_i -2 y_i y^j \partial_j ,\ \ \
\ee
we recognize $d,p_i,k_i,j_{ij}$ as the dS$_d$ transformations in the CFT-like basis.

Next are the components
\begin{align}
    {\cal M}_{IJ} =  Y_{I} \frac{\partial}{\partial Y^{J}} -  Y_{J} \frac{\partial}{\partial Y^{I}} \quad &\to \quad  \Omega_{IJ} \equiv \omega_I \partial_J  - \omega_J \partial_I \,.
\end{align}
These are the Killing vectors normal to the brane, they generate an ${\frak{so}}(N)$ algebra that acts linearly on the $\pi^I$'s, according to \eqref{eq:PiITransformation3},
\be  \delta_{\Omega_{JK}} \pi^I = \lp \pi_J \delta^I_K  - \pi_K \delta^I_J \rp \,.\label{sononbrefsymee}\ee

The rest of the Killing vectors are neither tangent nor normal. These are 
\begin{align}
    &{\cal M}_{0I} = -Y^{0} \frac{\partial}{\partial Y^{I}} -  Y_{I} \frac{\partial}{\partial Y^{0}} \quad \to \quad  \nn\\
    &~~ \, \Omega^{(0)}_I \equiv \omega_I \lp \frac{\tau^2 +y^2+1}{1-\rho^2} \partial_\tau + \frac{2 \tau  y^j\partial_j }{1-\rho^2}\rp - \frac{(\rho^2 -1 )(\tau^2-y^2-1)}{4 \tau} \partial_I + \frac{(\tau^2-y^2-1)}{2 \tau} \omega_I \omega^J \partial_J  \,,\nn\\ 
    &{\cal M}_{1I} = -Y^{0} \frac{\partial}{\partial Y^{I}} -  Y_{I} \frac{\partial}{\partial Y^{0}} \quad \to \quad \nn\\ 
    &~~ \, \Omega^{(1)}_I \equiv \omega_I \lp \frac{\tau^2 +y^2-1}{1-\rho^2} \partial_\tau + \frac{2 \tau y^j\partial_j }{1-\rho^2} \rp - \frac{(\rho^2 -1 )(\tau^2-y^2+1)}{4 \tau} \partial_I + \frac{(\tau^2-y^2+1)}{2 \tau} \omega_I \omega^J \partial_J \,,\nn\\ 
    &{\cal M}_{iI} =  Y_{i} \frac{\partial}{\partial Y^{I}} -  Y_{I} \frac{\partial}{\partial Y^{i}} \quad \to \quad \nn\\ 
    &~~ \,\Omega_{i I} \equiv  \frac{2 \omega_I}{\rho^2-1} \lp y_i \partial_\tau + \tau \partial_i \rp + \frac{1-\rho^2}{2 \tau} y_i \partial_I + \frac{y_i \omega_I}{\tau} \omega^J \partial_J\,.
\end{align}
For convenience, we use the following linear combinations,
\begin{align}
   \Omega^+_I &\equiv \Omega^{(0)}_I + \Omega^{(1)}_I = \frac{2 \omega_I  (\tau^2 +y^2)}{1-\rho^2} \partial_\tau + \frac{4 \omega_I  \tau }{1-\rho^2} y^j\partial_j  - (\tau^2-y^2) \llp \frac{(\rho^2-1)}{2 \tau} \delta^J_I - \frac{\omega_I \omega^J}{\tau}  \rrp \partial_J\,, \nn\\
    \Omega^-_I &\equiv \Omega^{(0)}_I - \Omega^{(1)}_I = \frac{2 \omega_I }{1 - \rho^2} \partial_\tau + \llp \frac{(\rho^2-1)}{2 \tau} \delta^J_I - \frac{\omega_I \omega^J}{\tau}  \rrp \partial_J \,.
 \end{align}
The transformation law of the $\pi^I$ field can be obtained by plugging these expressions into Eq.~\eqref{eq:PiITransformation}, and setting $\rho^2=\pi^2\equiv \delta_{IJ}\pi^I\pi^J$. These are
\begin{align}
 \delta_{\Omega^+_J} \pi^I &= (y^2 - \tau^2)\llp \frac{(\pi^2 -1)}{2 \tau } \delta^I_{\ J} - \frac{\pi^I \pi_J }{\tau}\rrp + (u^2 + y^2) \, \llp \frac{2  \pi_J \partial_\tau \pi^I}{\pi^2 -1}  \rrp + 4 \tau y^i \llp \frac{ \, \pi_J \partial_i \pi^I  }{\pi^2 -1 }  \rrp \,, \nn \\
    \delta_{\Omega^-_J} \pi^I &=  \llp \frac{(\pi^2 -1)}{2 \tau } \delta^I_{\ J} - \frac{\pi^I \pi_J }{\tau}\rrp + \llp \frac{2 \pi_J \partial_\tau \pi^I}{\pi^2 -1} \rrp \,, \nn\\
     \delta_{\Omega_{iJ}} \pi^I &= - y_i  \llp \frac{(\pi^2 -1)}{2 \tau } \delta^I_{\ J} - \frac{\pi^I \pi_J }{\tau}\rrp - y_i \llp  \frac{2 \, \pi_J    \partial_\tau \pi^I }{\pi^2 -1} \rrp - 2 \tau  \llp \frac{ \pi_J  \partial_i \pi^I }{\pi^2 -1} \rrp \,. \label{fillbrokentranse}
\end{align}
 These symmetries are spontaneously broken by the choice of the brane embedding.

\subsection{Scalar field action}

Comparing the explicit bulk metric \eqref{eq:MetricExample} to the general form \eqref{genwrpedmetrhree}, we easily read off the functional forms of the warp factor and the orthogonal metric, 
\be\label{eq:WandHIJExample}
e^{2\W} = \frac{L^2_D}{L^2_d} \lp \frac{1 - \rho^2}{1 + \rho^2 } \rp^2\,, \qquad  H_{IJ} = \frac{4 L^2_D }{(1 + \rho^2)^2} \, \delta_{IJ}= \frac{4 L^2_d \: e^{2 \W}}{(1 - \rho^2)^2} \delta_{IJ}\,.
\ee 
To find the multi-field action we plug these into Eqs.~\eqref{eq:CCintermsofPI} and \eqref{eq:EHintermsofPI} and set $\rho^2=\pi^2\equiv \delta_{IJ}\pi^I\pi^J$.   

At this point we will restrict to $d=4$.  We obtain
\begin{align} 
    \sqrt{- \bar g}  &= \sqrt{-   g} \frac{L^4_D}{L^4_4} \Bigg[  \frac{(1-\pi^2)^4}{(1+\pi^2)^4} +  \frac{2 L^2_4 \: (1-\pi^2)^2 }{(1 +\pi^2)^4} \pi^{I \mu} \pi_{I\mu} + \nonumber\\
     &~~~~~~~~~~~~~~~~ +  \frac{2 L^4_4  }{(1 + \pi^2)^4} \lp \pi^{I \mu} \pi_{I\mu} \pi^{J \nu} \pi_{J \nu} - 2 \pi^{I \mu}  \pi^{J \nu} \pi_{J\mu} \pi_{I \nu} \rp \Bigg] +{\cal O}\left(\nabla^6\right)\,,\label{eq:CCUpto4Deriv}\\
    \sqrt{-\bar g} \bar R &= \sqrt{-  g} \, \frac{L^2_D}{L^4_4}\Bigg[ 12  \frac{(1-\pi^2)^2}{(1+\pi^2)^2} + \frac{12  L^2_4 }{(1+\pi^2)^2} \pi^{I\mu} \pi_{I\mu} + \frac{96 L^2_4 \pi^I \pi^J \pi_I^{\mu} \pi_{J\mu}}{(1+\pi^2)^4}  + \nonumber\\
    &~~~~ - \frac{32 L^4_4}{(1+\pi^2)^2} \lp \frac{\pi_J \pi^J_{\ \mu\nu}}{(1-\pi^4)} + \frac{\pi^J_\mu \pi_{J\nu}}{(1-\pi^4)}+ \frac{4 (1+\pi^2) \pi^J \pi^K \pi_{J \mu} \pi_{K\nu}}{(1-\pi^4)^2}\rp \pi^{I\mu} \pi_I^\nu + \nonumber\\
    &~~~~ + \frac{32 L^4_4}{(1+\pi^2)^2} \lp \frac{\pi_J  \Box \pi^J}{(1-\pi^4)} + \frac{\pi^{J\nu} \pi_{J\nu}}{(1-\pi^4)}+ \frac{2(1-2\pi^2) \pi^J \pi^K \pi^\nu_{J} \pi_{K\nu}}{(1-\pi^4)^2} \rp \pi^{I\mu} \pi_{I\mu} \Bigg] +{\cal O}\left(\nabla^6\right)  \,,\label{eq:EHUpto4Deriv}
\end{align}
where we have used the shorthand notation $ \pi^I_\mu = \nabla_\mu \pi^I \, , \   \pi^I_{\mu\nu} = \nabla_\mu\nabla_\nu \pi^I $.
For brevity we have displayed only the terms with up to four derivatives, though ultimately we will also need the six derivative terms to find the quartic galileon, which can be obtained starting from the expressions in Eqs.~\eqref{eq:CCintermsofPI} and \eqref{eq:EHintermsofPI}.
The full theory is then obtained by substituting all this into \eqref{fullactioneqe}.

\subsection{Vacua}

 We first look for vacuum solutions of the total Lagrangian \eqref{fullactioneqe} that preserve the dS$_d$ symmetry on the brane.  These are the solutions for which $\pi^I=const.$, which we can find by extremizing the potential.  Using \eqref{eq:WandHIJExample} to evaluate the potential \eqref{genpotentiale}, we have
\begin{align}\label{eq:PotentialExample}
    V(\pi) &=  a_2 \frac{L^4_D}{L^4_4} \llp  \lp \frac{1-\pi^2}{1+\pi^2} \rp^4   -C \lp \frac{1-\pi^2}{1+\pi^2} \rp^2  \rrp  \,,
\end{align}
 where we have defined the dimensionless parameter 
 \be C \equiv (12 a_4) / (a_2 L^2_D)\, ,\label{C4def}\ee
 which will be convenient for characterizing the extrema.
The potential is a function only of $\pi\equiv\sqrt{\delta_{IJ}\pi^I\pi^J}$, exhibiting the $\frak{so}(N)$ symmetry \eqref{sononbrefsymee}. 

\begin{figure}
    \centering
    \includegraphics[width=0.8\textwidth]{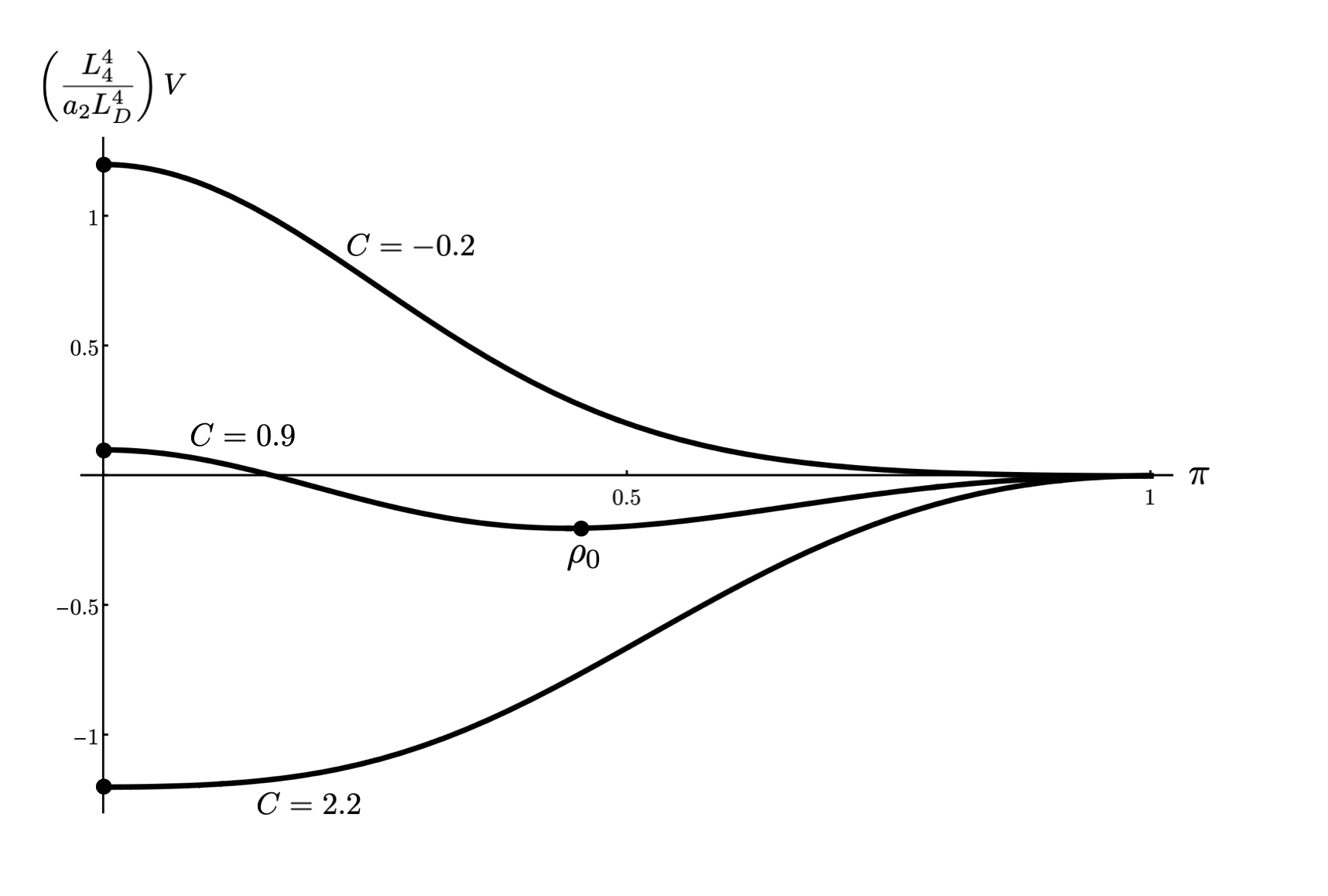}
    \caption{The scalar potential for representative choices of $C$.  The extrema are indicated by dots: If $C \leq 0$, the potential has a local maximum at $\pi=0$.  If $ 0 < C <2$, the potential has a local maximum at $\pi=0$ and a local minimum at $\pi =\rho_0>0$.  For  $C \geq 2$ there is a local minimum at $\pi =0$.}
    \label{fig:second}
\end{figure}

In the working range $\pi \in [ 0, 1)$, the potential goes to zero as $\pi\rightarrow 1$, and is smooth at the origin, approaching a constant as $\pi\rightarrow 0$ with vanishing derivative.  The existence of extrema within this range depends on the value of $C$.  In what follows we make the assumption that $a_2>0$.  (If $a_2<0$, we can take $a_4\rightarrow -a_4$ and pull an overall sign out of the action, and then all the statements about stability will simply be reversed, e.g. stable$\leftrightarrow$unstable, ghostly$\leftrightarrow$healthy, minimum$\leftrightarrow$maximum, etc.)
 \begin{itemize}
    \item  If $C \geq 2$, there is one extremum, a local minimum at $\pi = 0$.  
    \item If  $ 0 < C <2$, there are two extrema: a local maximum at $\pi=0$, and a non-trivial local minimum at $\pi=\rho_0$ where
    \be \rho_0 \equiv \frac{\sqrt{2} - \sqrt{C}}{\sqrt{2 - C}} .\label{nontrivialminneege}\ee
   The location of this minimum ranges over $(1,0)$ as $C$ ranges over $(0,2)$.
    \item  If $C\leq 0$, there is one extremum, a local maximum at $\pi=0$.
 \end{itemize}
 A plot of the potential with representative values of $C$ for each of these cases is shown in Fig.~\ref{fig:second}.

We see that there are two kinds of extrema, those at $\pi=0$ that preserve the $\frak{so}(N)$ symmetry and exist for any value of $C$, and those at $\pi=\rho_0$ that spontaneously break the $\frak{so}(N)$ symmetry and exist only in the range $0< C <2$.
In the remaining subsections we study perturbations around these two kinds of vacua.

\subsection{Expansion around $\frak{so}(N)$ preserving vacua}\label{sec:ExpansionAroundRhoZero}

We first consider the $\pi=0$ vacua, which realize the symmetry breaking pattern
\be \frak{so}(1,D) \rightarrow  \frak{so}(1,d)\oplus  \frak{so}(N) \,.\ee
To expand the action around the vacuum solution $\pi = 0$, we simply expand in powers of $\pi^I$. 
Apart from a constant term that does not contribute to the $\pi^I$ dynamics, the expansion of the cosmological constant and Einstein-Hilbert contribution up to fourth order in the fields, and to all orders in derivatives (which go up to order 6 at this order in the fields), are given by
\begin{align} 
    \sqrt{- \bar g}  &= \sqrt{- g} \frac{L^4_D}{L^4_4}  \Big[   2 L^2_4   \pi^{I \mu} \pi_{I\mu} -8\, \pi^2  \nn\\
    &~~~~~~~~~~~~~~~~+  2 L^4_4  \left( \pi^{I \mu} \pi_{I\mu} \pi^{J \nu} \pi_{J \nu} - 2 \pi^{I \mu}  \pi^{J \nu} \pi_{J\mu} \pi_{I \nu} \right)    -{12 } L^2_4\, \pi^2 \pi^{I \mu} \pi_{I\mu} +  {32 \,\pi^4}  \Big]+{\cal O}\left(\pi^6\right) \,,\\
    \sqrt{-\bar g} \bar R &= \sqrt{-  g} \, \frac{L^2_D}{L^4_4}\Big[ 12 L^2_4 \pi^{I \mu} \pi_{I\mu} -{48 \pi^2} \nn\\
    &~~~~~~~~~~~~~~~~+ 16 L^6_4 \pi_I^{\ \mu} \pi_J^{\ \nu} \lp \pi^{I \rho}_{\ \ \nu} \pi^J_{\ \rho \mu} - \pi^I_{\ \mu\nu}  \Box \pi^J\rp    \nonumber\\
    &~~~~~~~~~~~~~~~~ + 32 L^4_4   \left( \pi^{I\mu} \pi_{I\mu} \pi_J  \Box \pi^J -\pi^{I\mu} \pi_I^\nu \pi_J \pi^J_{\ \mu\nu} +  \pi^{I\mu} \pi_{I\mu} \pi^{J\nu} \pi_{J\nu} -\pi^{I\mu} \pi_I^\nu \pi^J_\mu \pi_{J\nu}\right) + \nonumber\\
    &~~~~~~~~~~~~~~~~ + 24  L^2_4 \left(  -  \pi^2 \pi^{I\mu} \pi_{I\mu} + 4 \pi^I \pi^J \pi_I^{\mu} \pi_{J\mu} \right)   + {96\, \pi^4}  \Big] +{\cal O}\left(\pi^6\right)\,,
\end{align}
where $\pi^I_\mu \equiv \nabla_\mu \pi^I \, , \ \    \pi^I_{\mu\nu} \equiv \nabla_\mu\nabla_\nu \pi^I$.

Collecting the quadratic terms of the full action \eqref{fullactioneqe}, we find
\be
{1\over \sqrt{-  g}} {\cal L}_{(2)} =   a_2\left(2-C\right)\frac{L^4_D}{L^2_4}  \left( - \pi^{I \mu} \pi_{I\mu} +\frac{4 }{L^2_4} \pi^2 \right) \,,\label{fuquasfdxralage}
\ee 
with $C$ as in \eqref{C4def}.  This is the sum of $N$ de Sitter galileons.   The mass of all the fields is fixed to be 
\be m^2 = 4 /L^2_4\,,\label{massvalue} \ee  
and the $\frak{so}(N)$ symmetry is manifest.
In the small field limit, the broken symmetry transformations \eqref{fillbrokentranse} become
 \begin{align}
   \delta_{\Omega^+_J} \pi^I &= - \frac{  y^2 - \tau^2}{2 \tau } \delta^I_{\ J}\,,  \nn \\
    \delta_{\Omega^-_J} \pi^I &=   - \frac{1}{2 \tau } \delta^I_{\ J} \,, \nn\\
    \delta_{\Omega_{iJ}} \pi^I &=   \frac{1}{2 \tau } y_i  \delta^I_{\ J}  \,.  \label{galishifgsyemee}
\end{align}
This is $N$ copies of the galileon shift symmetries on dS, or the $k=1$ shift symmetries in the language of \cite{Bonifacio:2018zex,Bonifacio:2021mrf}.  The mass \eqref{massvalue} is the $k=1$ shift symmetric value fixed by these transformations.  

The overall sign of the quadratic action \eqref{fuquasfdxralage} depends on the value of $C$:
\begin{itemize}
\item When $C< 2$, the mass term is tachyonic, since the $\pi=0$ extremum is a local maximum in this case.  The kinetic term, whose sign is fixed relative to the mass term by the shift symmetry \eqref{galishifgsyemee}, is positive.  Despite the fact that this is a local maximum of the potential, this is actually a healthy vacuum: the scalar is described by a unitary representation of the dS group (it is a member of the discrete series, see \cite{Boers:2013pba,Basile:2016aen,Sun:2021thf,Sengor:2022lyv,Sengor:2022kji,Enayati:2022hed,RiosFukelman:2023mgq,Schaub:2024rnl} for some recent reviews of dS representation theory).
\item When $C>2$,  the mass term is not tachyonic since the $\pi=0$ extremum is a local minimum in this case.  However the kinetic term, whose sign is still fixed relative to the mass term by the shift symmetry \eqref{galishifgsyemee}, is ghostly.  This is therefore an unstable vacuum, despite being at a local minimum of the potential.
\item When $C=2$, the kinetic term vanishes.  This is a point of strong coupling.  We can see this by looking at the quartic interaction terms: canonically normalizing the kinetic term via $\hat \pi \sim \left|2-C\right|^{1/2}a_2^{1/2}\frac{L^2_D}{L_4} \pi$,  the quartic interactions with the highest number of derivatives go like $\sim {1\over \Lambda^6} \nabla^6 \hat\pi^4$ with a strong coupling scale $\Lambda\sim {|2-C|^{1/3}\over C^{1/6}}\left( {a_2L_D^4\over L_4^6}\right)^{1\over 6} $, which goes to zero as $C\rightarrow 2$.
\end{itemize}

When $C=2$ the leading Lagrangian is quartic in the fields,
\begin{align}
    {1\over \sqrt{- g}}    {\cal L}_{(4)} &=  {8\over 3} a_2 L_D^4 L_4^2 \Bigg[    \pi_I^{\ \mu} \pi_J^{\ \nu} \left( \pi^{I \rho}_{\ \ \nu} \pi^J_{\ \rho \mu} - \pi^I_{\ \mu\nu}  \Box \pi^J\right)    \nonumber\\
    &~~~~ + {1\over L^2_4} \left(- 2\pi^{I\mu} \pi_I^\nu \pi_J \pi^J_{\ \mu\nu}  +2 \pi^{I\mu} \pi_{I\mu} \pi_J  \Box \pi^J - \frac12\pi^{I\mu} \pi_I^\nu \pi^J_\mu \pi_{J\nu}+\frac{5}{4}  \pi^{I\mu} \pi_{I\mu} \pi^{J\nu} \pi_{J\nu} \right)  \nonumber\\
    &~~~~ + {3\over L^4_4 } \left( \pi^2 \pi^{I \mu} \pi_{I\mu}  +2\pi^I \pi^J \pi_I^{\mu} \pi_{J\mu} \right)   -\frac{6 \pi^4}{L^6_4}  \Bigg]\,. \label{quarticlafgge} 
\end{align}
Since it is leading, \eqref{quarticlafgge} is still invariant under the lowest order transformation \eqref{galishifgsyemee}. It is a quartic multi-field de Sitter galileon, generalizing the quartic dS galileon of ~\cite{Goon:2011qf,Goon:2011uw,Burrage:2011bt} to the multi-field case\footnote{Specializing to $N=1$ and doing some integrations by parts, we can write it in the form 
\begin{align}
{3\over 8a_2 L_D^4 L_4^2  }  { 1\over \sqrt{- g}}   {\cal L}_{(4)}  &=\frac12 (\partial \pi)^2     \llp   ( \Box \pi)^2- \pi^{\mu\nu} \pi_{\mu\nu}   \rrp + {1\over L^2_4} \llp 3 (\partial \pi)^2  \pi  \Box \pi + \frac14 (\partial \pi)^4\rrp  +{9\over L_4^4}\pi^2 (\partial \pi)^2  -   {6\over L_4^6} {\pi^4} 
\end{align}
In this form, ${\cal L}_{(4)}$ is the same as the one given in~\cite{Burrage:2011bt}.}, and the flat space quartic multi-galileon of ~\cite{Hinterbichler:2010xn} to dS space.

\subsection{Expansion around the $\frak{so}(N)$ breaking vacua\label{sec:ExpansionSymmetryBreakingVacuum}}

We now look at the expansion of the theory around the non-zero minimum that occurs when $0<C<2$. 
This vacuum configuration spontaneously breaks the internal $\frak{so}(N)$ rotational symmetry among the $\pi^I$ fields to a $\frak{so}(N-1)$ subalgebra and, as such, the perturbed theory around it will not show explicitly the full $\frak{so}(N)$ symmetry but will only manifest the linearly realized $\frak{so}(N-1)$.  We thus have the symmetry breaking pattern
\be 
\frak{so}(1,D)  \rightarrow  \frak{so}(1,d)\oplus  \frak{so}(N) \rightarrow  \frak{so}(1,d)\oplus  \frak{so}(N-1) \,,
\ee
where the second step indicates the brane going from the trivial vacuum at $\pi=0$ to the non-trivial vacuum at $\pi=\rho_0$.
Since the second step involves breaking an $\frak{so}(N)$ internal symmetry down to $\frak{so}(N-1)$, there will be $N-1$ broken internal symmetry generators, and thus we can expect $N-1$ Goldstone bosons to appear.

We make the choice to align the vacuum configuration with the last scalar field $I=N$, thus we expand 
\be \pi^I = \pi^I_0 + \phi^I\,, \label{notexpdvaueeveee}\ee
with 
\be
\pi^I_0 = (0, \dots, \rho_0) = \rho_0 \, \delta^{I}_{N} \, ,
\ee
where $\rho_0$ is the non-zero minimum of the potential defined in \eqref{nontrivialminneege}.

It will be more convenient to express $C$ in terms of $\rz$ by inverting the relation \eqref{nontrivialminneege} between the two,
\be
C =   2 \frac{(1 - \rz^2)^2}{(1+\rz^2)^2} \,.
\ee 
With this, the total action \eqref{fullactioneqe} reads
\be\label{eq:TotalActionSSB}
{\cal L}_{tot} =  -a_2 \sqrt{-\bar g}  \lp 1- \frac{ L^2_D}{6} \frac{(1 - \rz^2)^2}{(1+\rz^2)^2}  \bar R\rp\,. 
\ee 

Expanding around the vacuum as in \eqref{notexpdvaueeveee}, the full quadratic action for the fluctuations reads
\be {1\over  \sqrt{-g} }{\cal L}_{2} =-16a_2 {L_D^4\over L_4^2}{\rz^2  (1-\rz^2)^2 \over  (1+\rz^2)^6} \left( -\nabla_\mu \phi_N \nabla^\mu \phi_N+{4\over L_4^2} \phi_N^2\right).\label{galmassphNee}\ee
We see a dS galileon action for the ``radial mode'' $\phi_N$, with the galileon mass \eqref{massvalue}, and a wrong sign kinetic term.  However, the remaining modes, the expected Goldstone modes, are missing.

The absence of the Goldstone modes in the quadratic action can be understood as resulting from a clash between the galileon symmetry and the broken internal symmetry.    Expanding the symmetry transformations \eqref{fillbrokentranse} around the $\rho_0$ vacuum, the transformations acting on the radial mode $\phi^N$ look just like those of \eqref{galishifgsyemee}, up to a constant rescaling  depending on $\rho_0$,
 \begin{align}
   \delta_{\Omega^+_J} \phi^N &= - \frac{  y^2 - \tau^2}{2 \tau } \left(1+\rho_0^2\right) \delta^I_{\ J}\,,  \nn \\
    \delta_{\Omega^-_J} \phi^N &=   - \frac{1}{2 \tau } \left(1+\rho_0^2\right) \delta^I_{\ J} \,, \nn\\
    \delta_{\Omega_{iJ}} \phi^N &=   \frac{1}{2 \tau } y_i  \left(1+\rho_0^2\right) \delta^I_{\ J}  \,.  \label{galishifgsyemee2}
  \end{align}
This fixes the galileon mass in \eqref{galmassphNee}.  The transformations acting on the Goldstone modes $\phi^{I\not=N}$ are similar, but with a slightly different constant rescaling depending on $\rho_0$,
 \begin{align}
     \delta_{\Omega^+_J} \phi^{I\not=N} &= - \frac{  y^2 - \tau^2}{2 \tau } \left(1-\rho_0^2\right) \delta^I_{\ J}\,,  \nn \\
    \delta_{\Omega^-_J} \phi^{I\not=N} &=   - \frac{1}{2 \tau } \left(1-\rho_0^2\right) \delta^I_{\ J} \,, \nn\\
    \delta_{\Omega_{iJ}} \phi^{I\not=N} &=   \frac{1}{2 \tau } y_i  \left(1-\rho_0^2\right) \delta^I_{\ J}  \,.  
\end{align}
If the Goldstones had a kinetic term, this symmetry would require the galileon value \eqref{massvalue} for the mass.  However, for the Goldstones there is another non-linearly realized symmetry: at zeroth order in the fields the broken component of \eqref{sononbrefsymee} gives a constant shift when acting on the Goldstones,
\bea && \delta_{\Omega_{JN}} \phi^I = -\rho_0 \delta^I_J\, ,\ \  \ I,J\not=N\, ,\nn\\ 
&& \delta_{\Omega_{JN}} \phi^N=0\, ,\ \ \ \ \  \ \  \ J\not=N \,.
\eea
The transformation on the Goldstones is the usual leading non-linear constant shift transformation of a Goldstone field and it requires that the mass vanish.  The only way to have this symmetry and the galileon symmetry at the same time, at leading order in the fields, is for the kinetic term to vanish. On the contrary, the radial mode $\phi^N$ does not transform under the shift symmetry, hence this mode is allowed to have a mass and a kinetic term.

However, the Goldstone modes do not vanish completely beyond linear order, and are present in the interactions.  For example, when
we expand Eq.~\eqref{eq:TotalActionSSB} up to third order in the field perturbations, we find the Goldstones:
\begin{align}
  {1\over   \sqrt{-g} }{\cal L}_{3} &= -a_2  \frac{L^4_D}{L^4_4} \lp \frac{-16 L^2_4  (1-\rz^2)^2}{(1+\rz^2)^6} \rp  \Bigg[ 
 \phi_N \lp \frac{16 \rz^3 (2 - 3 \rz^2 + \rz^4)}{L^2_4 (1+\rz^2)(1-\rz^2)^2} \phi^2_N - \frac{4(1+\rz^2)^2 }{L^2_4 }  \phi^I \phi_I \rp  + \nonumber\\
    &~~~~ + \phi_N \lp \frac{\rz(3 \rz^2 - 1)}{(1-\rz^2)} \phi^{I\mu} \phi_{I\mu}  -\frac{8 \rz^3 }{(1+\rz^2)} \phi_N^{\mu} \phi_{N\mu}  - 2 \rz \phi^I  \Box \phi_I \rp  \Bigg]\,.
\end{align}
We conclude, then, that the Goldstone modes are present but are infinitely strongly coupled\footnote{The reasoning for this interpretation is as follows: if we have a very small kinetic term, proportional, say, to some small quantity $\epsilon^2$, for a field that appears nonlinearly without dependence on $\epsilon$, then canonically normalizing the field will put powers of $1\over \epsilon$ in front of the fields in the interactions, and so the interaction strengths will go to infinity as $\epsilon\rightarrow 0$.  Thus we can interpret a vanishing kinetic term for a field that appears non-linearly as infinitely strong coupling.}. 

In a situation like this where there are fewer modes in the linear theory around some background as compared to the nonlinear theory, we say that there are Boulware-Deser ghosts with respect to the background.  The ghosts are the degrees of freedom in the full theory that do not appear in the linear theory.  This terminology comes from \cite{Boulware:1972yco}, where it was seen that a generic interacting theory of massive gravity, which has five degrees of freedom at the linear level around flat space, actually has six degrees of freedom non-linearly.   The extra modes are called ghosts because when expanding around a different background they tend to propagate, and typically with wrong signs for their kinetic terms.  For example, the ghost mode in massive gravity can be seen when expanding about spherically symmetric solutions sourced by a heavy mass \cite{Creminelli:2005qk,Deffayet:2005ys}, about which it has a wrong sign kinetic term.   (Unlike our Goldstones which are infinitely strongly coupled, the massive gravity ghost does not propagate on flat space because it is infinitely massive).

In our case, when $0<C<2$ we have the situation where there are two dS vacuum solutions.  One of them, the $\frak{so}(N)$ breaking solution, propagates only one degree of freedom, but we know that the full theory has $N$ degrees of freedom (thanks to the second order equations of motion ensured by the galileon structure), and so this vacuum has $N-1$ Boulware-Deser ghosts (in addition to the more traditional wrong-sign kinetic term ghost due to the wrong sign in \eqref{galmassphNee} for the remaining propagating mode).  There is, however, the other $\frak{so}(N)$ preserving vacuum around which all the modes propagate and there are no Boulware-Deser ghosts or wrong sign kinetic terms.   This demonstrates that the presence of a Boulware-Deser ghost, like the more traditional wrong-sign kinetic term ghost, is not a priori a fatal flaw but rather depends on the background and the structure of the rest of the theory: even if they are present around one background, there can be other backgrounds in which all the modes are healthy and not strongly coupled.   We have seen here an example where this happens without breaking dS symmetry in a theory containing only scalar fields.

\section{Conclusions}

Many well-known scalar field theories with higher-derivative interactions can be interpreted as effective theories describing the bending modes of branes embedded in higher-dimensional spacetimes.
The specific couplings that arise in these theories enable a variety of remarkable features, including screening mechanisms, self-accelerating solutions, and non-renormalization theorems, each of which carries deep implications for both gravitational and non-gravitational field theories in various contexts.
Given this richness, it is natural to further develop the probe brane construction and explore its consequences in broader contexts. 

In this work, we have used the probe brane construction to construct galileon-like theories in the case of higher co-dimension $N$ and allowing for both curved bulk and brane geometries, extending and bridging previous approaches that separately explored higher co-dimension embeddings in flat spacetime and co-dimension one scenarios in curved geometries.  We have illustrated the application of these results through a specific example: a four-dimensional de Sitter brane embedded in a $(4+N)$-dimensional de Sitter bulk.  
We have explicitly computed the Killing vectors of the bulk spacetime, which generate the symmetries of the $N$ scalar fields  $\pi^I$, leading to multi-field DBI galileon theories on dS$_4$ that non-linearly realize the symmetries of a $(4+N)$-dimensional de Sitter space broken down to the symmetries of dS$_4$ and an internal $\frak{so}(N)$.

We found that the potential of the resulting theory admits multiple dS$_4$ vacuum solutions: for some range of the parameters in front of the cosmological constant and Einstein-Hilbert terms on the brane, the theory admits both trivial $\pi = 0$ and non-trivial  $\pi \neq 0$ vacuum solutions.  
We then studied perturbations around the two possible vacuum solutions.  
When expanding around $\pi = 0$ solutions, the theory preserves the full $\frak{so}(N)$ symmetry and reduces to an action that is a multi-field generalization of the well-known de Sitter galileon theory.  
The broken dS symmetries fix the mass squared of the $N$ scalar fields to $m^2=4/L_4^2$, and we explicitly derived the quartic galileon Lagrangian $\mathcal{L}_{(4)}$ characteristic of these theories.  
On the other hand, when expanding the general action around the $\pi \neq 0$ vacuum, we find several interesting features.  
As expected, the rotational symmetry is spontaneously broken down to $ \frak{so}(N-1)$, resulting in $N-1$ Goldstone bosons.  
However, the latter are not only massless, but they also lack a kinetic term, while still appearing in interactions, implying that they are infinitely strongly coupled.
This result is further supported by analyzing the symmetries of the perturbation theory around this vacuum:  
the transformation laws of the scalar fields still contain the standard de Sitter galileon symmetries, but the Goldstone bosons also enjoy an additional constant shift symmetry, which is the non-linear realization of the original $\frak{so}(N)$ symmetry.  Due to the simultaneous presence of these two kinds of shift symmetry, there exists no second-order action that remains invariant, and consequently, the quadratic action for the Goldstone modes must vanish.  This gives an example of a theory with two dS vacua which propagate different numbers of degrees of freedom: what looks like Boulware-Deser ghosts around one vacuum (the missing Goldstone modes) become non-ghostly degrees of freedom around the other.

There are many other avenues of investigation that would be interesting to explore.  At tree level, these theories are dS examples of what in flat space have been called ``multi-$\rho$'' effective field theories \cite{Padilla:2016mno,Kampf:2021tbk}, which have mixed power counting and an intricate structure of soft limits.  It would be interesting to see how this, and the geometric soft theorems of \cite{Cheung:2021yog}, might extend to dS space, perhaps along the lines of \cite{Armstrong:2022vgl}.  Beyond tree level, the galileons and brane models have non-renormalization properties \cite{dePaulaNetto:2012hm,Brouzakis:2013lla,Goon:2020myi} that would be interesting to understand in these more general setups.  Going beyond the maximally symmetric vacuum solutions we have studied, it would be interesting to examine the fate of the vanishing Goldstone degrees of freedom around other solutions near the non-trivial vacuum, such as sourced spherical solutions (the $\frak{so}(N)$ multi-field  
galileons are known to develop issues such as negative gradients and superluminal propagation around such solutions \cite{Andrews:2010km,Goon:2010xh,Garcia-Saenz:2013gya}).
Finally, it might be interesting to see what can be said about the flavor kinematics duality of \cite{deNeeling:2022tsu} in the dS context.

{\bf Acknowledgments:}
A.G. is supported by funds provided by the Center for Particle Cosmology at the University of Pennsylvania.  KH acknowledges support from DOE grant DE-SC0009946. The work of M.T. is supported in part by the DOE (HEP) Award DE-SC0013528.


\appendix

\section{Higher orders in the derivative expansion of the action}\label{sec:CC}

To derive the derivative expansion of the action, it is convenient to introduce the notation
\be
\tilde {g}_{\mu\nu} \equiv  e^{2 \W} g_{\mu\nu} \,,  \qquad h_{\mu\nu} \equiv  H_{IJ} \TD_\mu \pi^I \TD_\nu \pi^J\,, \label{Hconfee}
\ee
where $\TD_\mu$ is the covariant derivative of $\tilde g_{\mu\nu}$.  With this, the induced metric \eqref{genwrpedmetrhree} can be written as 
\be
\bar g_{\mu\nu} = \tilde g_{\mu\nu} + h_{\mu\nu}\,,
\ee 
since $\partial_\mu \pi^I = \TD_\mu \pi^I$.
We can then expand in powers of $h_{\mu\nu}$ about the background $\tilde g_{\mu\nu}$. 
Since $\tilde g_{\mu\nu}$ does not depend on derivatives of $\pi^I$, this is then equivalent to a derivative expansion of the action, with each new power of $h_{\mu\nu}$ bringing in two more derivatives.  We perform a Weyl transformation at the end to express the results in terms of the brane metric $g_{\mu\nu}$ rather than $\tilde g_{\mu\nu}$.

Expanding the cosmological constant and Einstein-Hilbert contributions in powers of $h_{\mu\nu}$ is now an exercise identical to the one of linearizing gravity around a curved background. We will need the terms up to 6th order in derivatives, so we need to expand the cosmological constant, which comes with no derivatives to begin with, up to 3rd order in $h_{\mu\nu}$, and we need to expand the Einstein-Hilbert term, which comes with 2 derivatives to begin with, up to 2nd order in $h_{\mu\nu}$:
\begin{align}
    \sqrt{-\bar g} &\approx \sqrt{-{\tilde g}} \left[  1 + \frac{h}{2}+ \frac{1}{8} \lp  h^2 - 2 h^{\mu\nu} h_{\mu\nu} \rp + \frac{1}{48}\left( h^3-6 h h^{\mu\nu} h_{\mu\nu} +8h^{\mu}_{\ \nu} h^{\nu}_{\ \rho} h^{\rho}_{\ \mu}  \right)  \right]    + \mathcal{O}(h^4)\,, \label{eq:CCSecondOrder}\\
    \sqrt{- \bar g} \bar R &\approx \sqrt{-{\tilde g}} \Bigg( {\tilde R} - h^{\mu\nu} {\tilde G}_{\mu\nu} + \lp {\tilde R}_{\mu\nu} - \frac{1 }{4} {\tilde g}_{\mu\nu} \tilde R \rp \lp h^\mu_\rho h^{\rho \nu}   - \frac12 h h^{\mu\nu} \rp + \nonumber\\
    &~~~~~~~~~~~ - \frac14 \,   \tilde\nabla^\lambda h^{\mu\nu}  \tilde\nabla_\lambda h_{\mu\nu} + \frac12 \tilde\nabla_\rho  h^{\mu\nu}  \tilde\nabla_\mu h^\rho_\nu  -\frac12 \tilde\nabla_\mu   h \tilde\nabla_\nu h^{\mu\nu}  + \frac14 \tilde\nabla^\mu h  \tilde\nabla_\mu  h\Bigg) + \mathcal{O}(h^3) \,.\label{eq:EHSecondOrder}
\end{align}
In these expressions $\tilde R, \tilde G_{\mu\nu}, \tilde R_{\mu\nu}, \tilde\nabla_\mu$ are the Ricci scalar, Einstein tensor, Ricci tensor and covariant derivative, respectively, computed with the metric $\tilde g_{\mu\nu}$, which is also used to raise and lower all the indices.

Substituting in the expression \eqref{Hconfee} for $h_{\mu\nu}$ in terms of $\pi^I$, and performing the Weyl transformation to get from $\tilde g_{\mu\nu}$ to $g_{\mu\nu}$, we obtain
\begin{align}
         \sqrt{- g}  &= \sqrt{-  \hat f} e^{ d \W} \Bigg[ 1 + \frac{\hat H_{IJ}  }{2} \pi^{I \mu} \pi^J_\mu + \frac{\hat H_{IJ} \hat H_{KL}}{8} \lp \pi^{I\mu} \pi^J_{\ \mu} \pi^{K\nu} \pi^L_{\ \nu} - 2 \pi^{I\mu} \pi^{J\nu} \pi^{K}_{\ \mu} \pi^L_{\ \nu} \rp + \nonumber\\
        & + \frac{\hat{H}_{IJ} \hat{H}_{KL} \hat{H}_{MN}}{24} \lp 4 \pi^I_\mu \pi^J_\nu \pi^{K \rho} \pi^{L \nu} \pi^{M \mu} \pi^N_{\rho} - 2 \pi^{I \rho} \pi^J_{\ \rho} \pi^{K \mu} \pi^{L \nu} \pi^M_{\ \mu} \pi^N_{\ \nu} - \pi^{I \mu} \pi^J_{\ \mu} \pi^{K \nu} \pi^L_{\ \nu} \pi^{M \rho} \pi^N_{\ \rho}\rp \Bigg]  + \nonumber \\
        &~~~~+ \mathcal{O}(h^4)\,,\label{eq:CCintermsofPI}\\
        \sqrt{-\bar g} \bar R &= \sqrt{-  g} e^{(d-2) \W} \Bigg[  R - \lp  2 (d-1) {\Box} \W + (d-2)(d-1) \W^\lambda \W_\lambda  \rp  -  \hat H_{IJ}   G_{\mu\nu} \pi^{I\mu} \pi^{J\nu}+ \nonumber\\
        &~~~~- (d-2)  \hat H_{IJ} \pi^{I\mu} \pi^{J\nu} \lp \W_\mu \W_\nu - \W_{\mu\nu}\rp - \frac{(d-2)}{2}  \hat H_{IJ} \pi^{I\mu} \pi^{J}_{\ \mu}  \lp (d-3) \W^\lambda \W_\lambda + 2  \Box \W \rp +\nonumber\\ 
        &~~~~ + \hat H_{IJ} \hat H_{KL}\lp \hat \pi^{I\mu} \pi^J_{\ \rho} \pi^{K \nu} \pi^{L \rho}  - \frac12 \pi^{I \rho} \pi^J_{\ \rho} \pi^{K \mu} \pi^{L \nu} \rp \lp \hat R_{\mu\nu} - \frac14 \hat f_{\mu\nu} \hat R \rp + \nonumber\\
        &~~~~ +  (d-2) \hat H_{IJ} \hat H_{KL}\lp \hat \pi^{I\mu} \pi^J_{\ \rho} \pi^{K \nu} \pi^{L \rho}  - \frac12 \pi^{I \rho} \pi^J_{\ \rho} \pi^{K \mu} \pi^{L \nu} \rp \lp \W_\mu \W_\nu - \W_{\mu\nu} \rp + \nonumber\\
        &~~~~ + \hat H_{IJ} \hat H_{KL}\lp \hat \pi^{I\mu} \pi^{J\nu} \pi^K_{\ \mu} \pi^L_{\ \nu} - \frac12  \pi^{I\mu} \pi^{J}_{\ \mu} \pi^{K \nu} \pi^L_{\ \nu}   \rp \lp \frac{(d-1) (d-2) }{4}  \W^\lambda \W_\lambda  + \frac{(d+1)}{2}  \Box \W \rp + \nonumber\\
        &~~~~ +  \frac{(2-d)}{2} \hat H_{IJ} \pi^I_\rho \pi^{J\mu} \,  \W^\rho \, \llp    \pi^{K\nu}  \pi^L_{\ \nu}  \nabla_\mu \hat H_{KL}  +  2 \hat H_{KL} \pi^{K\nu} \pi^L_{\ \mu \nu} \rrp  \nonumber\\
        &~~~~+\frac12 \lp \nabla_\rho \hat H_{IJ} \nabla_\mu  \hat H_{KL} \rp \pi^{I\nu} \pi^{K \rho} \llp \pi^{J \mu} \pi^L_{\ \nu} - \pi^{L \mu} \pi^J_{\ \nu} \rrp + \nonumber\\
        &~~~~ -\frac14 \lp \nabla^\rho \hat H_{IJ} \nabla_\rho  \hat H_{KL} \rp \pi^{I\mu} \pi^{K \nu} \llp \pi^J_{\ \nu} \pi^L_{\ \mu} - \pi^J_\mu \pi^L_{\ \nu} \rrp  + \nonumber\\
        &~~~~ + \lp \hat H_{IJ} \nabla_\rho \hat H_{KL}\rp  \pi^I_{\mu\nu} \pi^{K \nu} \llp \pi^{J \rho} \pi^{L \mu} - \pi^{J \mu} \pi^{L \rho} \rrp + \nonumber\\
        &~~~~+ \frac12 \lp \hat H_{IJ} \nabla_\rho \hat H_{KL}\rp \pi^I_{\ \nu} \llp \pi^{J \nu \rho} \pi^{K \mu} \pi^L_{\ \mu} - \pi^J_{\ \nu} \pi^{L \rho}  \Box \pi^K \rrp + \nonumber\\
        &~~~~+ \lp \hat H_{IJ} \hat H_{KL} \rp \pi^{I \mu} \pi^{K \nu} \llp \pi^{J \rho}_{\ \ \ \nu} \pi^L_{\ \rho \mu }  - \pi^J_{\ \mu\nu}   \Box \pi^L \rrp\Bigg]+ \mathcal{O}(h^3)\,. \label{eq:EHintermsofPI}
\end{align}
In these expressions $ R,  G_{\mu\nu},  R_{\mu\nu}, \nabla$ are the Ricci scalar, Einstein tensor, Ricci tensor and covariant derivative, respectively, computed with the brane metric $g_{\mu\nu}$, which is also used to raise and lower all indices, and $d$ is the brane dimension. Additionally, we have used the following notations: 
\begin{align}
& \hat H_{IJ} \equiv e^{-2 \W} H_{IJ},\ \ \     \pi^I_\mu = \nabla_\mu \pi^I \, , \ \    \pi^I_{\mu\nu} = \nabla_\mu\nabla_\nu \pi^I \, ,\ \ \  \W_\mu = \nabla_\mu \W \, ,  \ \ \  \W_{\mu\nu} = \nabla_\mu \nabla_\nu \W\,.
\end{align}

\renewcommand{\em}{}
\bibliographystyle{utphys}
\addcontentsline{toc}{section}{References}
\bibliography{ManyGalileonsCurvedSpace_arxiv}

\end{document}